\documentclass[epj]{svjour}
\usepackage{graphicx,color}
\begin{document}
  \newcommand {\nc} {\newcommand}
  \nc {\beq} {\begin{eqnarray}}
  \nc {\eeq} {\nonumber \end{eqnarray}}
  \nc {\eeqn}[1] {\label {#1} \end{eqnarray}}
  \nc {\eol} {\nonumber \\}
  \nc {\eoln}[1] {\label {#1} \\}
  \nc {\ve} [1] {\mbox{\boldmath $#1$}}
  \nc {\ves} [1] {\mbox{\boldmath ${\scriptstyle #1}$}}
  \nc {\mrm} [1] {\mathrm{#1}}
  \nc {\half} {\mbox{$\frac{1}{2}$}}
  \nc {\thal} {\mbox{$\frac{3}{2}$}}
  \nc {\fial} {\mbox{$\frac{5}{2}$}}
  \nc {\la} {\mbox{$\langle$}}
  \nc {\ra} {\mbox{$\rangle$}}
  \nc {\etal} {\emph{et al.}}
  \nc {\eq} [1] {(\ref{#1})}
  \nc {\Eq} [1] {Eq.~(\ref{#1})}
  \nc {\Ref} [1] {Ref.~\cite{#1}}
  \nc {\Refc} [2] {Refs.~\cite[#1]{#2}}
  \nc {\Sec} [1] {Sec.~\ref{#1}}
  \nc {\chap} [1] {Chapter~\ref{#1}}
  \nc {\anx} [1] {Appendix~\ref{#1}}
  \nc {\tbl} [1] {Table~\ref{#1}}
  \nc {\Fig} [1] {Fig.~\ref{#1}}
  \nc {\ex} [1] {$^{#1}$}
  \nc {\Sch} {Schr\"odinger }
  \nc {\flim} [2] {\mathop{\longrightarrow}\limits_{{#1}\rightarrow{#2}}}
  \nc {\IR} [1]{\textcolor{red}{#1}}
  \nc {\IB} [1]{\textcolor{blue}{#1}}
  \nc{\IG}[1]{\textcolor{bobcatgreen}{#1}}
  \nc{\pderiv}[2]{\cfrac{\partial #1}{\partial #2}}
  \nc{\deriv}[2]{\cfrac{d#1}{d#2}}

\title{Study of Cluster Structures in Nuclei through the Ratio Method}
\subtitle{A Tribute to Mahir Hussein}
\author{Pierre Capel\inst{1,2}\thanks{pcapel@uni-mainz.de} \and Ronald C. Johnson\inst{3}\thanks{r.johnson@surrey.ac.uk} \and Filomena M. Nunes\inst{4}\thanks{nunes@frib.msu.edu}
}                     
%
%

\institute{Institut f\"ur Kernphysik, Johannes Gutenberg-Universit\"at Mainz, D-55099 Mainz, Germany \and 
Physique Nucl\'eaire et Physique Quantique (CP 229), Universit\'e libre de Bruxelles (ULB), B-1050 Brussels, Belgium \and
Department of Physics, University of Surrey, Guildford GU2 7XH, United Kingdom \and
National Superconducting Cyclotron Laboratory and Department of Physics and Astronomy, Michigan State University, East Lansing, MI 48824, USA}
\date{Received: date / Revised version: date}
%
\abstract{
For one-neutron halo nuclei, the cross section for elastic scattering and breakup at intermediate energy exhibit similar angular dependences.
The Recoil Excitation and Breakup (REB) model of reactions elegantly explains this feature.
It also leads to the idea of a new reaction observable to study the structure of loosely-bound nuclear systems: the Ratio.
This observable consists of the ratio of angular distributions for different reaction channels, viz. elastic scattering and breakup, which cancels most of the dependence on the reaction mechanism; in particular it is insensitive to the choice of optical potentials that simulate the projectile-target interaction.
This new observable is very sensitive to the structure of the projectile.
In this article, we review the Ratio Method and its extension to low beam energies and proton-halo nuclei.
} 
\maketitle
\section{Introduction}
\label{intro}

Since their development in the mid-80s, Radioactive-Ion Beams (RIBs) have provided a unique way to explore the nuclear chart away from stability.
This technical breakthrough has led to the discovery of unexpected structures.
In particular, some nuclei close to the neutron dripline have been found to exhibit a matter radius much larger than their isobars \cite{Tan85b}, which contradicts the usual description of the nucleus as a tight pileup of nucleons.
Further analyses have shown that this unusually large size is due to the lose binding of one or two valence nucleons, which can then exhibit a high probability of presence at a large distance from the other nucleons.
Such nuclei are usually seen as a compact core, which contains most of the nucleons, around which one or two neutrons form a sort of diffuse halo \cite{Tan96}, hence their name: \emph{halo nuclei} \cite{HJ87}.
The best known halo nuclei are $^{11}$Be, with a one-neutron halo structure, and $^{11}$Li, which is a two-neutron halo nucleus.
On the proton-rich side of the nuclear chart, proton halos are also possible, though less probable.
For example, $^8$B most likely exhibits a one-proton halo.

Being located far from the bottom of the valley of stability, halo nuclei exhibit very short lifetimes, which make them difficult to study.
Often reactions are the only way to infer information about their structure.
Various experiments have been devised to better understand the origin of these exotic nuclei \cite{Tan96}: elastic scattering \cite{Dip10,Dip12}, transfer \cite{Sch12,Sch13,Wim18}, knockout \cite{Aum00}, and breakup \cite{Pal03,Fuk04}.
In parallel, significant efforts have been put by theorists to develop models of these reactions in order to reliably infer nuclear-structure information from these measurements \cite{BC12}.
The Continuum Discretised Coupled Channel model (CDCC) \cite{Aus87} initially developed to describe deuteron-induced reactions, has been successfully extended to analyse the elastic-scattering and breakup of halo nuclei.
It is also included in a Coupled-Channel Born Approximation (CCBA) of transfer reactions \cite{GCM14,Joh14}.
More often, the Adiabatic Distorted Wave Approximation (ADWA) is used to describe transfer  \cite{JS70,JT74,TJ20}.
At sufficiently high beam energy, the eikonal approximation may be used to simplify the  reaction model \cite{Glauber}.
This approximation is mostly used in the analysis of knockout experiments \cite{HT03} or breakup reactions at intermediate energies, e.g. within the Dynamical Eikonal Approximation (DEA) \cite{BCG05,GBC06}.

Throughout his professional life Mahir S.\ Hussein (1944--2019)  contributed significantly to this exciting field of research, working on different fronts. In particular, he developed models for various types of reactions that improved our understanding of the reaction mechanism and thereby helped us infer more accurate nuclear-structure information from experiment \cite{HM84,HM85,HLN06,GCL11,TCH17,Pot17}.
While playing such an active role within the community of nuclear-reaction theorists, Mahir was also very supportive towards the youngsters, providing them with enlightening new perspectives on their projects, which helped them progress in their work.
His deep warm voice and welcoming smile, the lively twinkle at the corner of his eye and his benevolent behaviour will be missed among his friends and colleagues.

The work discussed in this paper uses breakup reactions, including those driven by the Coulomb interaction.
Traditionally, experimentalists have used subtraction techniques to remove the nuclear component from the breakup cross section measured on a heavy target, in an attempt to obtain purely Coulombic cross sections.
Typically, the breakup cross section is also measured on a light target where the process is nuclear driven, and then the cross section is scaled and subtracted from the data on the heavy target.
Mahir Hussein and collaborators in \Ref{HLN06} expose the limitations of this technique and advocated for an approach that includes both nuclear and  Coulomb interactions in the analysis, avoiding subtraction altogether.
The method we discuss in the present contribution avoids this subtraction issue while providing direct access to nuc\-lear-structure information about the projectile.

\begin{figure}
  \begin{center}
    \includegraphics[width=\columnwidth]{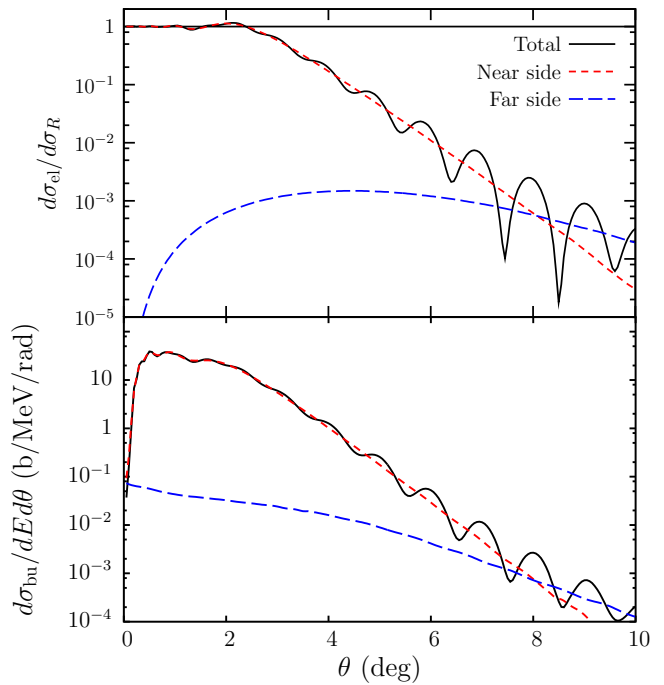}
\end{center}
  \caption{\label{f1} Theoretical analysis of the collision of $^{11}$Be on Pb at 69~MeV/nucleon.
The elastic-scattering cross section plotted as a ratio to Rutherford (top) and breakup angular distribution (bottom) are decomposed into their near (red short-dashed lines) and far (blue long-dashed lines) sides \cite{CHB10}.
Reprinted from \Ref{CHB10} with permission from Elsevier.}
\end{figure}

During a visit to Brussels, Mahir suggested an extension of the near/far decomposition of the elastic-scattering cross section \cite{HM84} to the angular distribution for the break\-up of one-neutron halo nuclei.
This idea led to a nice piece of work in which it was realised that in their collision with a target, halo nuclei are scattered similarly whether they remain bound, i.e. when they are elastically scattered, or when the halo dissociates from the core during the breakup of the projectile \cite{CHB10}.
This result is illustrated in \Fig{f1}, which presents DEA calculations for the collision of $^{11}$Be on Pb at 69~MeV/nucleon.
The top panel displays the elastic-scattering cross section plotted as a ratio to Rutherford, while the bottom panel contains the breakup cross section of $^{11}$Be into $^{10}$Be and a neutron at a relative energy $E=0.5$~MeV as a function of the scattering angle of the core-neutron centre of mass.
We observe that the sudden drop of these angular distributions (at $\theta\sim2^\circ$) and the oscillatory pattern at larger angles are very similar.
Note also that the near (red short-dashed lines) and far (blue long-dashed lines) contributions to those cross sections also look alike.

Interestingly, this result can be understood qualitatively using the Recoil Excitation and Breakup (REB) description of the collision \cite{JAT97,Joh98}.
In that model, the excitation of the halo nucleus, possibly leading to its breakup into the core and valence neutron, takes place through the recoil of the core due to its interaction with the target, while the neutron remains mostly unaffected and accordingly acts as a spectator.
This model leads to an elegant factorisation of the cross sections for both the elastic scattering and the breakup into the cross section for the elastic scattering of a pointlike projectile times a form factor that accounts for the actual extension of the halo.
Not only does the REB explain the results obtained in collaboration with Mahir Hussein, but it also suggests a new reaction observable: the \emph{Ratio} of selected cross sections for breakup and elastic channels \cite{CJN11}.
Following the REB prediction, this observable should be independent of the reaction process and hence be very sensitive to the projectile structure.
In this article, we review the idea of this Ratio Method.
In \Sec{theory}, we present the few-body reaction models considered here.
We explain how this new reaction observable is derived and show, in \Sec{ratio}, how the REB prediction compares to accurate reaction calculations.
We then summarise in Secs.~\ref{E20} and \ref{phalo} the extension of this idea beyond the range of validity of the REB, viz. to low-energy collisions and proton-halo nuclei, respectively.
In \Sec{numerical}, we suggest that the Ratio Method could be applied more widely than to single-nucleon halo nuclei.
Our conclusions are drawn in \Sec{conclusion}.

\section{Theoretical framework}\label{theory}

\subsection{Three-body model of collision}

To describe theoretically the collision involving a one-nucleon halo nucleus, we consider the usual three-body model of reactions \cite{BC12}.
The exotic projectile $P$ is described as a two-cluster system: a spinless core $c$, which contains most of the nucleons to which a valence nucleon N is loosely bound.
The internal structure of this system is captured by the single-particle Hamiltonian
\beq
H_0=-\frac{\hbar^2}{2\mu}\Delta+V_{c\rm N}(r),
\eeqn{e1}
where $\vec{r}$ is the relative coordinate between the core and the nucleon, $\mu$ is their reduced mass, and $V_{c\rm N}$ is an effective potential, whose parameters are adjusted to reproduce known structure observables of the nucleus, such as its one-nucleon separation energy and the spin and parity of its low-lying states.
The physical states of the projectile are then described by the eigenstates of $H_0$.
In the partial wave $ljm$, where the total angular momentum $j$ is obtained as the coupling of the $c$-N orbital angular momentum $l$ with the nucleon $\half$ spin, and where $m$ is the projection of $j$, we thus have
\beq
H_0\ \phi_{ljm}(E,\ve{r})=E\ \phi_{ljm}(E,\ve{r}),
\eeqn{e1a}
with the wave functions decomposing into a radial part $u_{lj}$ and a spin-angular part ${\cal Y}_{ljm}$:
\beq
\phi_{ljm}(\ve{r})=\frac{1}{r} u_{lj}(E,r)\ {\cal Y}_{ljm}(\Omega_r).
\eeqn{e1b}
The negative-energy eigenstates are discrete.
They correspond to bound $c$-N levels and are identified with the additional quantum number $n$ of nodes in the radial wave function $u_{nlj}$.
They are normed to unity and behave asymptotically as
\beq
u_{nlj}(E_{nlj},r)\flim{r}{\infty} {\cal C}_{nlj}\ W_{-\eta,l+1/2}(2\kappa_{nlj}r),
\eeqn{e1c}
where ${\cal C}_{nlj}$ is the asymptotic normalisation constant (ANC) of the bound state, $\kappa_{nlj}=\sqrt{2\mu \left|E_{nlj}\right|/\hbar^2}$, and $W_{-\eta,l+1/2}$ is the Whittaker function \cite{AS70}, which depends on the $c$-N Sommerfeld parameter $\eta=Z_cZ_{\rm N}\mu e^2/4\pi\epsilon_0\hbar^2 \kappa_{nlj}$, with $Z_c$ and $Z_{\rm N}$, the atomic number of the core and the valence nucleon, respectively.

The positive-energy eigenstates of $H_0$ describe the $c$-N continuum, i.e., the broken up projectile.
Their radial part is normalised according to
\beq
\lefteqn{u_{lj}(E,r)\flim{r}{\infty}\sqrt{\frac{2\mu}{\pi\hbar^2 k}}\left[\cos\delta_{lj}(E) F_l(\eta,kr)\right.}\hspace{4cm}\nonumber\\
& + &\left. \sin\delta_{lj}(E) G_l(\eta,kr)\right],
\eeqn{e1d}
where 
$\delta_{lj}$ is the phaseshift induced by $V_{c\rm N}$ in the partial wave $lj$, $k=\sqrt{2\mu E/\hbar^2}$, and $F_l$ and $G_l$ are the regular and irregular Coulomb wave functions, repsectively \cite{AS70}, which depend on the Sommerfeld parameter $\eta=Z_cZ_{\rm N}\mu e^2/4\pi\epsilon_0\hbar^2 k$.

As usual in few-body models of reactions with loosely-bound projectiles, the internal structure of the target $T$ is neglected and its interactions with the projectile constituents, $c$ and N, are optical potentials $V_{cT}$ and $V_{{\rm N}T}$, respectively, chosen from the literature \cite{BC12}.
The Hamiltonian that describes this three-body model of the reaction reads
\beq
H=-\frac{\hbar^2}{2\mu_{PT}}\Delta_R+H_0+V_{cT}(R_{cT})+V_{{\rm N}T}(R_{{\rm N}T}),
\eeqn{e2}
where $\ve{R}$ is the coordinate of the projectile centre of mass relative to the target, $\mu_{PT}$ is the $P$-$T$ reduced mass, and $R_{cT}$ and $R_{{\rm N}T}$ are the $c$-$T$ and N-$T$ distances, respectively.
Within this three-body framework, the study of the $P$-$T$ collision corresponds to solving the \Sch equation
\beq
H\ \Psi(\ve{R},\ve{r})=E_T\ \Psi(\ve{R},\ve{r}),
\eeqn{e3}
with the initial condition that the projectile, in its ground state $\phi_{n_0l_0j_0m_0}$, impinges on the target
\beq
\Psi^{(m_0)}(\ve{R},\ve{r})\flim{Z}{-\infty}e^{iKZ+\cdots}\phi_{n_0l_0j_0m_0}(\ve{r}),
\eeqn{e4}
where we have chosen the $Z$ axis along the incoming beam, and where the initial wave number $K$ is related to the total energy $E_T=\hbar^2K^2/2\mu_{PT}+E_{n_0l_0j_0}$.

One of the most accurate ways to solve the \Sch \Eq{e3} is to expand the three-body wave function $\Psi$ upon the projectile eigenstates $\phi_{ljm}$ and solve the corresponding coupled equations.
To take into account the channels in which the halo nucleon dissociates from the core, it is necessary to include a description of the projectile continuum.
This can be done by discretising it into small energy bins.
The corresponding model is known as the Continuum-Discretised Coupled Channel method, or CDCC \cite{Aus87}.
A publicly available code to solve the CDCC equations is {\sc fresco} \cite{fresco}.

At intermediate beam energies, i.e. above 40~MeV/nu\-cleon, the \Sch \Eq{e3} can be reliably solved using the Dynamical Eikonal Approximation (DEA) \cite{BCG05,GBC06}.
This approximation is built on the eikonal description of the collision \cite{Glauber}, but does not include the adiabatic treatment of the projectile dynamics.
The DEA leads to excellent agreement with data for both one-neutron and one-proton halo projectiles \cite{GBC06,GCB07}.
At sufficiently high beam energy, it also compares very well with CDCC \cite{CEN12}.

\subsection{Recoil Excitation and Breakup}\label{REB}

The striking similarity between the elastic-scattering cross section and the angular distribution for the breakup of $^{11}$Be on Pb at 69~MeV/nucleon illustrated in \Fig{f1} can be easily explained within a simpler model of the reaction than the computationally intensive CDCC or DEA.
The Recoil Excitation and Breakup model (REB) developed to describe reactions involving one-neutron halo nuclei incorporates the fact that the excitation of the projectile, potentially leading to its breakup, is mostly due to the tidal force experienced by the nucleus during the collision \cite{JAT97,Joh98}.
This tidal force appears because the core and valence neutron have a non-zero spatial separation in the projectile and do not interact in the same way with the target.
In intermediate-energy Coulomb dominated reactions, the core is mostly repulsed by the target, while the neutron can be seen as a spectator, its interaction with the target being small.
In this simple picture, excitation and breakup of the projectile during the reaction result from the recoil of the core.

This picture leads to a significant simplification of the \Sch \Eq{e3} through two approximations \cite{JAT97,Joh98}: (i) treating the projectile dynamics adiabatically and (ii) assuming  $V_{{\rm N}T}$ is negligible relative to $V_{cT}$.
The first one amounts to neglecting the excitation energy of the projectile compared to the beam energy and hence replacing $H_0$, the projectile internal Hamiltonian \eq{e1}, by a constant.
Choosing that constant equal to the projectile ground state energy $E_{n_0j_0l_0}$ enables us to satisfy the initial condition \eq{e4}.
The second one removes $V_{{\rm N}T}$ from the three-body Hamiltonian $H$ \eq{e2}.
With these two approximations, the resulting equation has no dynamic dependence on the $c$-N relative coordinate $\ve{r}$ (only parametric dependence).
An analytic factorization can then be found and the cross section for the elastic scattering of the projectile reads \cite{JAT97,Joh98}
\beq
\left( \frac{d \sigma}{d \Omega} \right)_{\rm el} = \left| F_{0,0}(\ve Q)\right|^2 \;  \left( \frac{d \sigma}{d \Omega} \right)_{\rm pt},
\eeqn{e5}
where $(d \sigma/d \Omega)_{\rm pt}$ is the elastic-scattering cross section obtained for a pointlike projectile of mass $\mu_{PT}$ scattered by $V_{cT}$, and the form factor $F_{0,0}$ accounts for the actual extension of the projectile halo:
\beq
\left|F_{0,0}(\ve Q)\right|^2=\frac{1}{2j_0+1}\sum_{m_0}\left|\int|\phi_{l_0j_0m_0}(\ve{r})|^2 e^{i\ve{Q\cdot r}}d\ve{r}\right|^2,
\eeqn{e6}
where $\ve{Q}= \frac{m_{\rm N}}{m_c+m_{\rm N}} (K\ve{\widehat{Z}} - \ve K')$ is proportional to the momentum transfered during the collision between the initial $K\ve{\widehat{Z}}$ and final $\ve K'$ $P$-$T$ momenta.
It relates to the scattering angle through
\beq
Q=2\frac{m_{\rm N}}{m_c+m_{\rm N}} K \sin(\theta/2).
\eeqn{e6a}

The REB thus enables us to separate the nuclear structure of the projectile from the reaction process, which is dominated by the $c$-$T$ interaction.
Interestingly, this can be extended to other reaction channels.
Following a similar idea, the cross section for the inelastic scattering to its bound state $i>0$ reads \cite{proc97,Joh98}
\beq
\left( \frac{d \sigma_i}{d \Omega} \right)_{\rm inel} = | F_{i,0}(\ve Q)|^2 \;  \left( \frac{d \sigma}{d \Omega} \right)_{\rm pt}.
\eeqn{e7}
with the form factor
\beq
\lefteqn{|F_{i,0}(\ve Q)|^2}\nonumber\\
&=& \frac{1}{2j_0+1}\sum_{m_0}\sum_{m_i}\left| \int\phi_{n_il_ij_im_i}(\ve{r}) \phi_{l_0j_0m_0}(\ve{r})
e^{i\ve{Q\cdot r}}d\ve{r}\right|^2.
\eeqn{e8}
Similarly, the breakup cross section at energy $E$ in the $c$-N continuum expressed as a function of the scattering angle $\Omega$ of the $c$-N centre of mass in the $P$-$T$ restframe reads
\beq
\left( \frac{d \sigma}{dE d \Omega} \right)_{\rm bu} = |F_{E,0}(\ve Q)|^2 \; \left(\frac{d \sigma}{d \Omega} \right)_{\rm pt},
\eeqn{e9}
using the form factor
\beq
\lefteqn{|F_{E,0}(\ve Q)|^2}\nonumber\\
&=& \frac{1}{2j_0+1}\sum_{m_0}\sum_{ljm} \left| \int\phi_{ljm}(E,\ve{r}) \phi_{l_0j_0m_0}(\ve{r})
e^{i\ve{Q\cdot r}}d\ve{r}\right|^2.
\eeqn{e10}
In both cases, we obtain a factorisation of the cross section into a cross section computed for the pointlike projectile times a form factor that accounts for the projectile's structure.
Note that in Eqs.~\eq{e8} and \eq{e10}, the different initial and final states of the projectile appear.
Interestingly, the cross section for the pointlike projectile is identical in all three expressions \eq{e5}, \eq{e7}, and \eq{e9}.
This explains the results of \Ref{CHB10} shown in \Fig{f1}: the projectile is scattered similarly by the target whether it stays in its ground state or if it is excited in another state, or even if it is broken up.
Besides explaining in simpler terms the results of dynamical calculations for the collision of $^{11}$Be, the REB suggests a new way to study the structure of loosely-bound nuclei through reactions: the Ratio Method \cite{CJN11,CJN13}.

\subsection{The Ratio Method}

The Ratio Method exploits the fact that the cross section for the pointlike projectile is identical in the REB cross sections \eq{e5}, \eq{e7}, and \eq{e9}. 
Therefore, taking the ratio of these angular distributions will cancel their dependence on the reaction mechanism, leaving a simple ratio of form factors, which depend only on $H_0$ eigenstates, hence producing a reaction observable highly sensitive to the projectile structure.
At least this is what the REB predicts.
Of course, the ratio of any linear combinations of the cross sections \eq{e5}, \eq{e7}, or \eq{e9} will also remove their dependency on $(d\sigma/d\Omega)_{\rm pt}$.
The detailed study presented in \Ref{CJN13}, and fruitful discussions with experimentalists, have shown that the best combination from a practical standpoint is that of the \emph{summed ratio}: 
\beq
{\cal R}_{\rm sum}(E,\ve Q) &=& \frac{(d\sigma/dEd\Omega)_{\rm bu}}{(d\sigma/d\Omega)_{\rm sum}}\label{e11a}\\
 &\stackrel{\rm (REB)}{=}&\left|F_{E,0}(\ve Q)\right|^2,
\eeqn{e11}
where the summed cross section corresponds to the sum of all elastic and inelastic processes
\beq
\lefteqn{\left(\frac{d\sigma}{d\Omega}\right)_{\rm sum}}\nonumber\\
 &=&\left(\frac{d\sigma}{d\Omega}\right)_{\rm el}
+\sum_{i>0}\left(\frac{d\sigma_i}{d\Omega}\right)_{\rm inel}+\int \left(\frac{d\sigma}{dEd\Omega}\right)_{\rm bu} dE.
\eeqn{e12} 

\begin{figure}
  \begin{center}
    \includegraphics[width=\columnwidth]{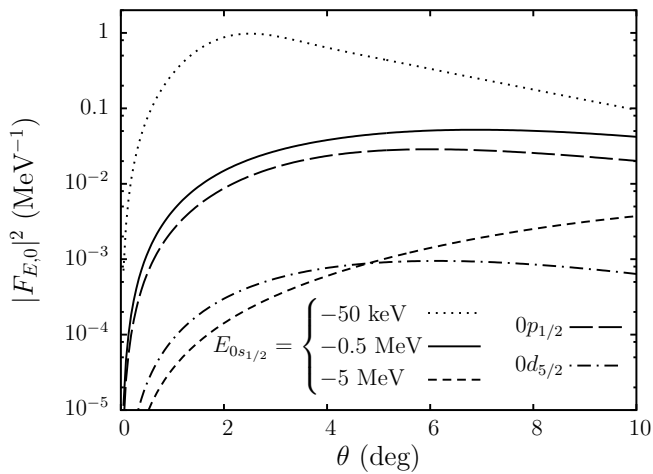}
\end{center}
  \caption{\label{f2} Sensitivity of the REB form factor $|F_{E,0}(\ve Q)|^2$ \eq{e10} to the projectile structure: the one-neutron separation energy in its ground state and the orbital in which the halo neutron is bound \cite{CJN11}.
Reprinted from \Ref{CJN11} with permission from Elsevier.}
\end{figure}

In the REB approximation, this summed ratio \eq{e11a} should be equal to the form factor $|F_{E,0}(\ve Q)|^2$ \eq{e10}.
Besides removing all dependence on the reaction process, this ratio should also be quite sensitive to the projectile structure.
To illustrate this, we produce in \Fig{f2} REB form factors obtained for a realistic $^{11}$Be projectile, viz. with a neutron bound to the $^{10}$Be core by 0.5~MeV in an $s_{1/2}$ orbital (solid line).
When that binding energy is reduced, resp. augmented, by a factor 10, both the shape and the magnitude of the form factor change significantly (dotted and short-dashed lines).
Similar changes are observed when the valence neutron is bound in a $p$ (long-dashed line) or a $d$ (dash-dotted line) state \cite{CJN11,CJN13}.
This shows that, if confirmed, the Ratio Method provides a very sensitive probe of the halo structure.
Because the changes observed in \Fig{f2} scale over different orders of magnitude, an actual ratio of cross sections that follows roughly the REB prediction \eq{e11} will provide more information than the cross sections for each individual reaction, from which it is built.

This idea, initially suggested in \Ref{CJN11} was analysed in detail in \Ref{CJN13} by confronting the REB prediction with actual dynamical calculations.
The encouraging results gathered in that analysis lead to the idea that the method might be extended outside the range of validity of the REB, viz. for low-energy reactions \cite{CCN16} and to one-proton halo nuclei \cite{YCP19}.
In the following sections, we summarise these different studies and present their major outcomes.

\section{Test of the idea for collisions involving $^{11}$Be at 70~MeV/nucleon}\label{ratio}

To test the idea of the Ratio Method, it is simplest to perform realistic calculations of reactions and compare their outcome for the ratio \eq{e11a} to the form factor predicted by the REB \eq{e10}.
For such a check, we have considered in Refs.~\cite{CJN11,CJN13} the collision of $^{11}$Be, the archetypical one-neutron halo nucleus, on Pb and C at about 70~MeV/nu\-cleon.
These are the experimental conditions under which the breakup channel has been measured at RIKEN by Fukuda \etal\ \cite{Fuk04}.
The reaction model used in that study is the DEA.
It provides the angular distributions needed to compute the summed cross section \eq{e12}, i.e. the elastic- and inelastic-scattering cross sections and the angular distributions for the breakup channel, and leads to an excellent agreement with the RIKEN data for different breakup observables on both targets \cite{GBC06,CPH18}.

\begin{figure*}
  \begin{center}
    \includegraphics[width=\columnwidth]{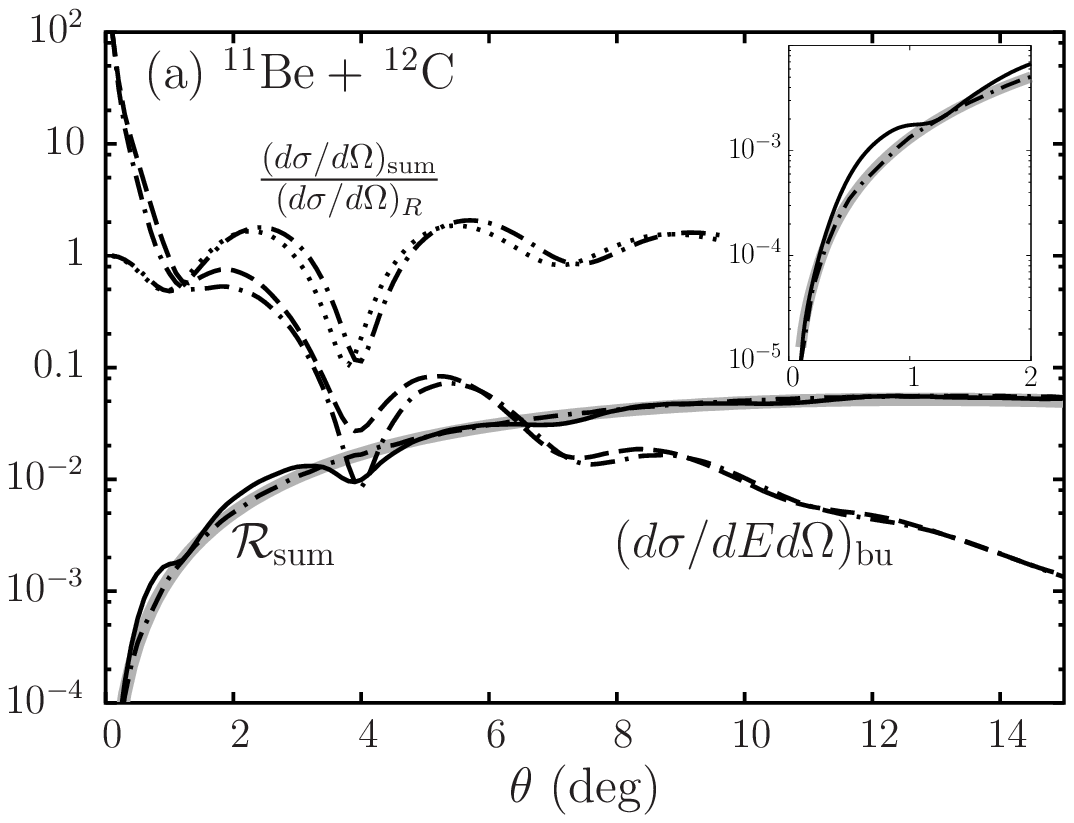}
    \includegraphics[width=\columnwidth]{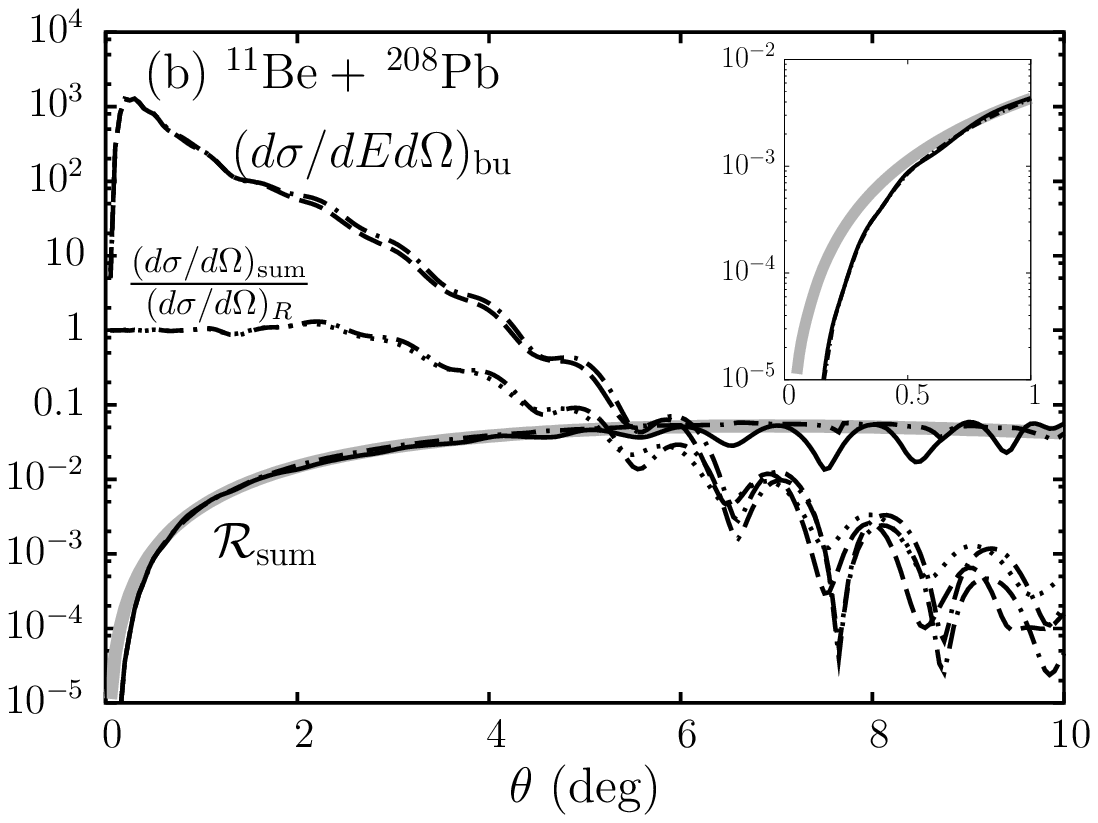}
\end{center}
  \caption{\label{f3} Analysis of the ratio method for the collisions of $^{11}$Be on (a) $^{12}$C at 67~MeV/nucleon  and (b) $^{208}$Pb at 69~MeV/nucleon.
  The ratio \eq{e11a} obtained from a DEA calculation of the reaction (solid black line) is compared to the REB prediction $\left|F_{E,0}(\vec{Q})\right|^2$ (thick gray line) alongside the summed cross section shown as a ratio to Rutherford (dotted line) and the breakup angular distribution (dashed line).
DEA calculations performed  without n-$T$ interaction are shown as dash-dotted lines \cite{CJN13}.
Reprinted figures with permission from \Ref{CJN13} Copyright (2013) by the American Physical Society.}
\end{figure*}

The results of this comparison are illustrated in \Fig{f3} for (a) the carbon target at 67~MeV/nucleon and (b) the lead target at 69~MeV/nucleon \cite{CJN13}.
Each panel depicts, as a function of the scattering angle of the projectile centre of mass, the summed cross section \eq{e12} as a ratio to Rutherford (dotted lines), the breakup angular distribution \eq{e9} (expressed in b/MeV\,sr) at a $^{10}$Be-n continuum energy of $E=0.1$~MeV (dashed lines), and their ratio ${\cal R}_{\rm sum}$ \eq{e11a} in units MeV$^{-1}$ (solid black lines).
The REB prediction of that ratio $|F_{E,0}(\ve Q)|^2$ \eq{e10} is plotted as the thick grey lines.

We first point out that our calculations confirm the results obtained in \Ref{CHB10}: the summed and breakup cross sections exhibit very similar patterns.
They oscillate nearly in phase and with similar magnitude.
Accordingly when we take their ratio most of their angular dependence cancels, which leads to a rather smooth curve in excellent agreement with the form factor predicted by the REB. 
Interestingly, this happens for both targets despite the very different reaction mechanisms: on $^{12}$C, the reaction is mostly dominated by the nuclear interaction and the breakup cross section remains small; whereas on $^{208}$Pb, the reaction is Coulomb dominated and the breakup cross section is large.
The Ratio \eq{e11a} therefore removes most of the sensitivity of the cross sections to the reaction process, leading to an observable that depends nearly exclusively on the projectile structure.
This agreement is not perfect, however.
On both targets we observe remnant oscillations in the DEA ratio, and at very forward angles on $^{208}$Pb, the REB form factor overestimates the DEA result.

The remnant oscillations in the DEA ratio are observed where the summed and breakup cross sections exhibit the most ample oscillations; at forward angles on $^{12}$C and beyond $5^\circ$ on $^{208}$Pb.
They are due to the slight shift that exists between the two angular distributions.
To understand the reason for that shift, we repeat the DEA calculations switching off the n-$T$ interaction (dash dotted lines in \Fig{f3}).
The remnant oscillations in that calculation vanish nearly completely, which suggests that this slight discrepancy between the dynamical calculations and the REB prediction is due to the hypothesis made in the latter that $V_{{\rm n}T}=0$.
The actual n-$T$ interaction slightly kicks the halo neutron affecting differently the angular distribution in the different reaction channels (elastic and inelastic scatterings and breakup) \cite{CJN13}.

The overestimation of the realistic ratio compared to the REB prediction is not due to $V_{{\rm n}T}$, as can be seen from the inset in \Fig{f3}(b), where the full dynamical calculation and the ratio obtained setting $V_{{\rm n}T}=0$ are superimposed.
Interestingly, this flaw is not observed on the carbon target.
We understand this issue as resulting from the other hypothesis made within the REB, viz. the adiabatic approximation.
That approximation is valid only for short collision times, i.e. when the interactions between the projectile and the target are short ranged.
While this makes sense for nuclear-dominated reactions, it is less valid for Coulomb breakup and leads to the divergence of the breakup cross section at forward angles \cite{GBC06} and hence a larger ratio than that obtained in the dynamical calculation \cite{CJN13}.

\begin{figure}[b]
  \begin{center}
    \includegraphics[width=\columnwidth]{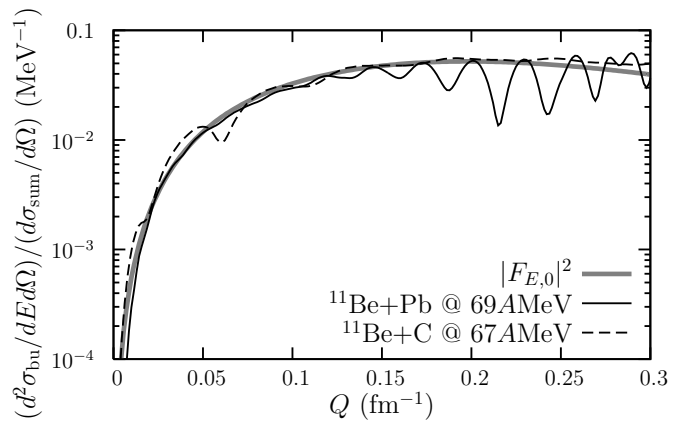}
\end{center}
  \caption{\label{f4}  Comparison of the ratio ${\cal R}_{\rm sum}$ \eq{e11a} obtained on two different targets $^{12}$C and $^{208}$Pb.
  Despite the very different reaction dynamics, both ratios are in good agreement with one another and with the REB prediction \eq{e11} \cite{CJN11}.
Reprinted from \Ref{CJN11} with permission from Elsevier.}
\end{figure}

These two issues remain small, and we can see from Figs.~\ref{f2} and \ref{f3} that the Ratio provides an observable much more sensitive to the projectile structure than individual cross sections.
The independence of the ratio from the reaction mechanism is illustrated in \Fig{f4}, where the DEA ratios obtained for the collision of $^{11}$Be on $^{12}$C at 67~MeV/nucleon (dashed line) and on $^{208}$Pb at 69~MeV/nu\-cleon (solid line) are compared to one another.
To remove the inherent difference in the beam energy and target mass, they are displayed as a function of $Q$ \eq{e6a}.
But for the aforementioned small remnant oscillations, both calculations lead to ratios very similar to one another, despite being driven by very different reaction mechanisms.
As mentioned above, these dynamical ratios are in good agreement with the REB prediction, confirming that this observable contains detailed information about the projectile structure \cite{CJN11,CJN13}.

The excellent results obtained in this analysis suggest that the Ratio is a reliable observable to study the internal structure of one-neutron halo nuclei at intermediate energy.
Since the approximations made to derive the REB lead to small effects on this observable, it is interesting to study the extension of this method to cases in which these approximations are less reliable, viz. for collision at lower beam energy \cite{CCN16} (see \Sec{E20}) and to one-proton halo nuclei \cite{YCP19} (see \Sec{phalo}).
The former extension would enable us to measure the ratio at larger scattering angles, and hence obtain a finer angular precision.
The latter would provide a reliable tool to study the structure of nuclei at or close to the proton dripline through reactions.

\section{Extension of the ratio to low energy (20~MeV/nucleon)}\label{E20}

\begin{figure*}
  \begin{center}
    \includegraphics[width=\columnwidth]{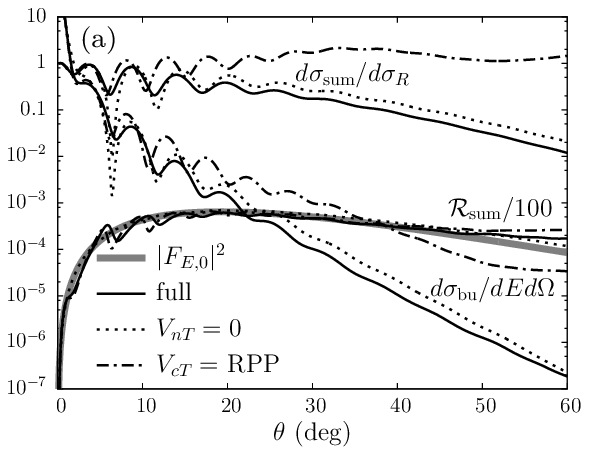}
    \includegraphics[width=\columnwidth]{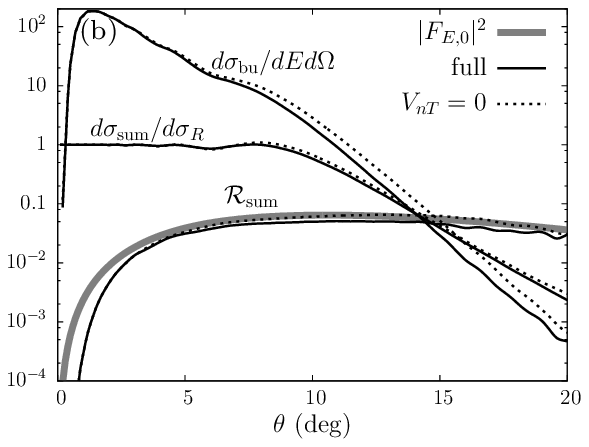}
\end{center}
  \caption{\label{f5} Analysis of the ratio method extended to low beam energy considering $^{11}$Be impinging at 20~MeV/nucleon on (a) $^{12}$C  and (b) $^{208}$Pb.
The solid black lines show the results of fully dynamical calculations of the reaction (CDCC on $^{12}$C and DEA on $^{208}$Pb).
The REB prediction $\left|F_{E,0}(\ve{Q}\right|^2$ for the ratio \eq{e11a} is shown by the thick gray line.
The dotted lines show the results of the dynamical calculations when $V_{{\rm n}T}$ is set to 0.
On the $^{12}$C target calculations using another $c$-$T$ interaction are shown with the dash-dotted lines \cite{CCN16}.
Reprinted figures with permission from \Ref{CCN16}. Copyright (2016) by the American Physical Society.}
\end{figure*}

In this subsequent study of the Ratio Method, we have looked at its potential use at lower beam energy, moving down to 20~MeV/nucleon \cite{CCN16}.
As in the previous section, we have compared fully dynamical calculations of reactions involving $^{11}$Be to the prediction of the REB for different target choices.
Because 20~MeV/nucleon is below the range of validity of the DEA on light targets, CDCC \cite{Aus87} has been used for the calculations on $^{12}$C, through its implementation in the code {\sc fresco} \cite{fresco}. 
On the heavy targets, we have used a correction to the DEA, which enables us to reach these beam energies for Coulomb-dominated reactions \cite{FOC14}.
In \Fig{f5}, we show the results obtained on (a) $^{12}$C and (b) $^{208}$Pb \cite{CCN16}.
As in \Fig{f3}, we display the summed cross sections \eq{e12} as ratios to Rutherford, the breakup angular distributions obtained at the $^{10}$Be-n continuum energy $E=125$~keV (expressed in b/MeV\,sr), and their ratio in MeV$^{-1}$ (note that on $^{12}$C, the ratio is divided by 100 for readibility).
In addition to the full dynamical calculation, which includes both $c$-$T$ and n-$T$ optical potentials (solid lines), we have also performed calculations without the n-$T$ interaction, to test that REB approximation (dotted lines).
On the $^{12}$C target [\Fig{f5}(a)], we have used an alternative $c$-$T$ optical potential to test the independence of the Ratio to that model input (RPP, dash-dotted lines).
The REB form factor is shown as the thick grey line.

On $^{12}$C, we observe that the results are very similar to those obtained at 67~MeV/nucleon [see \Fig{f3}(a)]: the ratio obtained from the CDCC calculations is in excellent agreement with the REB prediction \eq{e9}, but for small remnant oscillations.
These oscillations disappear when $V_{{\rm n}T}$ is set to zero.
The calculations performed with the alternative $^{10}$Be-$^{12}$C interaction show significant differences in the individual cross sections; at large angle, they can differ from the original calculation by two orders of magnitude.
Yet, their ratio is nearly superimposed with the ``full'' calculation, confirming the strong independence of this observable to the reaction mechanism, and that the Ratio exhibits very little dependence on this choice of inputs for the calculations.
This is very useful since optical potentials are usually difficult to constrain and are a dominant source of systematic uncertainty in reaction calculations \cite{CGB04}.
Although modern statistical tools to quantify uncertainties in reactions are being developed \cite{lovell2018,king2019,catacora2019}, they have so far been restricted to the parametric uncertainties in the nucleon-nucleus optical potentials. An extension of statistical studies to include nucleus-nucleus optical potentials would be very useful.

On the heavier target $^{208}$Pb, the results are not as convincing [see \Fig{f5}(b)].
Although we observe less remnant oscillations than on $^{12}$C---mostly because the cross sections for the different processes on the heavy target exhibit a smoother  angular dependence---, we see that the agreement with the REB prediction is less good than on $^{12}$C and at higher beam energy (see \Fig{f3}).
Most importantly, at forward angle, the issue stressed in \Sec{ratio} about the REB relying on the adiabatic approximation is amplified, which is expected for collisions taking place at lower energy.
In addition, at larger angles, the REB prediction overestimates the DEA ratio.
This is due to the presence of the n-$T$ interaction, which reduces the breakup cross section [compare the dotted and solid lines in \Fig{f5}(b)].

\begin{figure*}
  \begin{center}
    \includegraphics[width=\columnwidth]{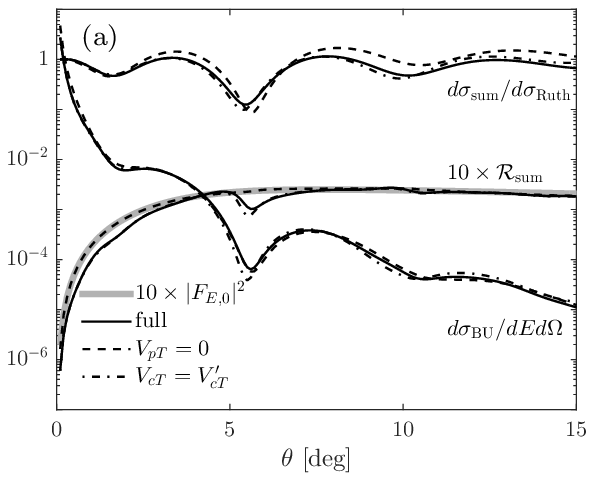}
    \includegraphics[width=\columnwidth]{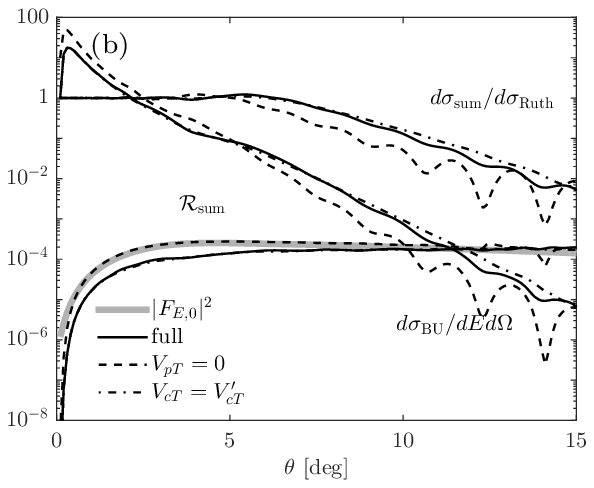}
\end{center}
  \caption{\label{f6} Extension of the ratio method to proton halos.
The collision of $^{8}$B at 44~MeV/nucleon on (a) $^{12}$C  and (b) $^{208}$Pb have been calculated with the DEA \cite{GCB07,YCP19}.
The solid black lines show the results of fully dynamical calculations, while the dashed lines correspond to calculations without the p-$T$ interaction.
Results using a different $V_{cT}$ are shown using dash-dotted lines.
The REB prediction $\left|F_{E,0}(\ve{Q}\right|^2$ for the ratio \eq{e11a} is shown by the thick gray line \cite{YCP19}.
\copyright IOP Publishing. Reproduced with permission from \Ref{YCP19}. All rights reserved.}
\end{figure*}

This analysis confirms the interest of the Ratio Method to study the structure of one-neutron halo nuclei and shows that it can also be used at beam energies down to 20~MeV/ nucleon with light targets \cite{CCN16}.
The nuclear-structure content of the ratio is similar to that at higher energy, its potential use at low beam energy broadens the range of RIB facilities where it could be experimentally implemented.
Although the Ratio Method removes most of the dependence on the reaction process, and hence should lead to equal results on different targets, reactions on light targets lead to better predictions by the REB at low beam energy.  It turns out that the strong Coulomb field generated by  heavy targets makes the adiabatic approximation, which is used in the REB, less reliable.

\section{Extension of the ratio to proton halos}\label{phalo}

Since the results summarised in the previous section illustrate that the Ratio Method can be used outside of the exact range of validity of the REB, upon which it is founded, it is prudent to check the applicability of the Ratio Method to study the structure of loosely-bound proton-rich nuclei such as proton halos \cite{YCP19}.
In those cases, the existence of a long-range Coulomb term in the interaction between the valence nucleon and the target is an additional challenge to the REB.
We have initiated this study considering a $^8$B projectile, whose very low $^7$Be-p threshold [$S_{\rm p}(^8{\rm B})=137$~keV] makes it the archetypical one-proton halo nucleus.
We have used the DEA, which has been shown to provide excellent agreement with experiment for this nucleus at beam energies between 44 and 83~MeV/nucleon \cite{GCB07}.
The results are illustrated in \Fig{f6} for collisions at 44~MeV/nucleon on (a) $^{12}$C and (b) $^{208}$Pb.
They include full DEA calculations (solid lines), a DEA calculation in which the p-$T$ optical potential is set to zero to estimate the influence of that interaction on the results ($V_{{\rm p}T}=0$, dashed lines), and another full calculation with a different choice of $c$-$T$ interaction to check the sensitivity of the Ratio ${\cal R}_{\rm sum}$ to that input ($V_{cT}=V'_{cT}$, dash-dotted lines).
In each case, the REB form factor $|F_{E,0}(\ve Q)|^2$ \eq{e10} is shown as a thick grey line.

As observed for one-neutron halo nuclei \cite{CHB10}, the angular distributions in the different reaction channels exhibit similar behaviour: same oscillatory pattern and, in the case of the lead target, a shoulder at $\theta\approx5^\circ$.
Accordingly, the ratio of the breakup to the summed cross sections removes most of that angular dependence.
However, as in the case of the low-energy collisions studied in \Ref{CCN16} and summarised in \Sec{E20}, we observe remnant oscillations in the DEA ratio on $^{12}$C and a clear overestimation of the DEA calculation by the REB ratio on the heavy target.
Interestingly, in this case, both issues are due to the presence in the realistic reaction calculation of the p-$T$ interaction: even on lead, switching off $V_{{\rm p}T}$ leads to a near-perfect agreement between the REB prediction and the dynamical calculations.
Whereas in the former case, the issue is due once more to the small shift induced by $V_{{\rm p}T}$ in the different cross sections, in the latter case, the issue is more subtle.
At forward angle, neglecting the p-$T$ interaction leads to a \emph{larger} breakup cross section.
This increase occurs because the repulsive Coulomb interaction between the halo proton and the target in the full calculation reduces the tidal force that leads to the breakup of the nucleus \cite{YCP19}.
That interaction thus changes significantly the reaction dynamics assumed in the REB, where the excitation and the breakup of the projectile occur mostly through the recoil of the core, and where the valence nucleon is seen as a spectator.
This result suggests that including this interaction, e.g. as a small perturbation, would improve the REB prediction, and accordingly lead to a better expression of the form factor to which experimental data could be compared.

\begin{figure*}
  \begin{center}
    \includegraphics[width=\textwidth]{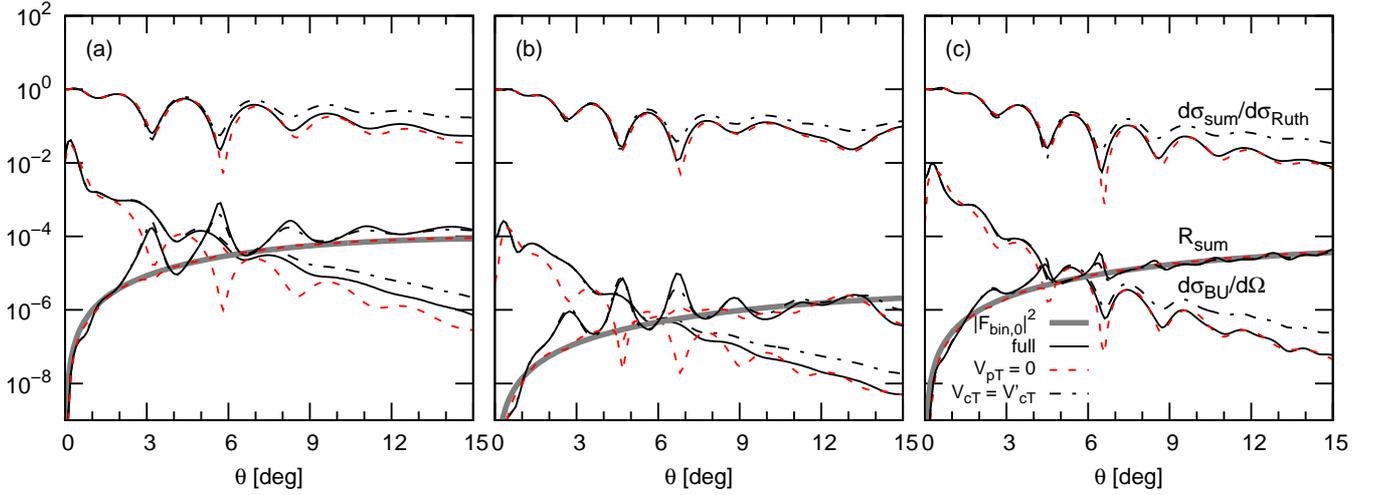}
\end{center}
  \caption{\label{f7} Analysis of the ratio \eq{e11a} for collision of proton-rich nuclei at 60~MeV/nucleon on $^{12}$C: (a) $^{17}$F, (b) $^{25}$Al, and (c) $^{27}$P \cite{YCP19}.
\copyright IOP Publishing. Reproduced with permission from \Ref{YCP19}. All rights reserved.}
\end{figure*}

In \Ref{YCP19}, a detailed analysis of the sensitivity of the Ratio observable to the projectile structure has shown that the agreement between DEA calculations and the REB prediction \eq{e11} deteriorates when the one-proton separation energy increases and when the valence proton is bound in an orbital with a larger orbital angular momentum.
The best cases are $s$ or $p$-wave states bound by less than 1~MeV, viz. nuclei which are  most likely to present a one-proton halo.
This is illustrated in \Fig{f7}, where DEA calculations performed for the collision of several projectiles on $^{12}$C (the most effective target for the Ratio Method)  at 60~MeV/nucleon are presented: (a) $^{17}$F, (b) $^{25}$Al, and (c) $^{27}$P.
The nucleus $^{17}$F has a clear $^{16}$O-p structure \cite{SBI00,Hag10}.
It is bound by only 0.6~MeV and its $\fial^+$ ground state is dominated by a $0d_{5/2}$ proton bound to $^{16}$O in its $0^+$ ground state \cite{Hag10}.
The second nucleus, $^{25}$Al has a one-proton separation energy of 2.3~MeV and the structure of its $\fial^+$ ground state is also dominated by a $0d_{5/2}$ proton.
Finally, $^{27}$P, has a $\half^+$ ground state that can be seen as a $1s_{1/2}$ proton bound by 0.87~MeV to its $^{26}$Si core.

In \Fig{f7}, the solid black lines correspond to full DEA calculations.
The red dashed lines show what happens when the p-$T$ interaction is set to zero, and the black dash-dotted lines, how the full calculation changes when a different $c$-$T$ optical potential is chosen.
For both $d$-bound nuclei, i.e. (a) $^{17}$F and (b) $^{25}$Al, we observe that contrary to the results of \Ref{CHB10}, the angular distributions for the different processes are no longer similar: the summed and breakup cross sections are nearly in opposition of phase with maxima in the former being located at the minima of the latter, and vice versa.
This leads to a DEA ratio with huge oscillations, which cannot be directly compared to the REB prediction.
In the case of $^{17}$F, this dissimilarity in the angular distributions is due solely to the p-$T$ optical potential.
When that interaction is switched off (see red dashed lines), the cross sections are perfectly in phase, leading to a smooth ratio, in excellent agreement with the REB form factor.
Note that the difference appears mostly in the breakup observable, indicating that for this nucleus, $V_{{\rm p}T}$ plays a significant role in the reaction, and cannot be realistically neglected, as hypothesised in the REB (see \Sec{REB} and Refs.~\cite{JAT97,Joh98}).
In the case of $^{25}$Al, on the contrary, forcing $V_{{\rm p}T}=0$ is not sufficient to provide a smooth ratio; even without p-$T$ interaction, the DEA ratio exhibits remnant oscillations.

The calculations involving $^{27}$P [see \Fig{f7}(c)] show a much more convincing case for the extension of the ratio method to proton-rich nuclei.
As observed for one-neutron halo nuclei \cite{CHB10}, the angular distributions for elastic scattering and breakup exhibit strong similarities.
Accordingly, the full DEA ratio merely oscillates around the REB prediction, and when $V_{{\rm p}T}$ is set to zero, the agreement with that form factor is nearly perfect.

The differences observed between all three cases can be easily understood from the systematic analyses of the Ratio presented in Refs.~\cite{CJN13,YCP19}.
In these articles, it has been shown that the Ratio Method works best for nuclei with a valence nucleon loosely bound to the core in a state with a low orbital angular momentum.
This explains why the REB form factor is in good agreement with the DEA ratio for $^{27}$P, while the idea does not seem to work for  nuclei with a  ground state described as a $d$-wave proton bound to the core.
This result suggests that, on the proton-rich side of the nuclear chart, the Ratio Method, in its original idea, can be used to study the structure of spatially extended loosely-bound structures such as proton halo nuclei.

\section{A numerical ratio method}\label{numerical}

So far we have focused on the ratio method resulting from a factorization of the elastic and breakup cross sections in the REB approximation. 
We have thus determined the accuracy of the method based on the agreement with the analytic form factor obtained when such a factorization occurs. We have shown that in several cases this factorization is not perfect due to the adiabatic assumption included in the derivation. However, both DEA and CDCC are reaction models that treat the dynamics accurately and therefore can generate a reliable ratio to compare to experiment.
Such a numerical approach does not have the simplicity of the original Ratio Method \cite{CJN11,CJN13}.
It does not enable the direct comparison of reaction measurements to the form factor \eq{e10} easily obtained from the projectile wave functions and it requires running computationally intensive codes.
Nevertheless, performing the ratio of  cross sections predicted with state-of-the-art reaction theories for elastic and breakup will enable meaningful comparisons with the corresponding experimental ratio, with the advantage that  the ambiguities related to the interactions with the target are essentially removed.

To illustrate this we focus on the $^{17}$F or $^{25}$Al  ratio shown in \Fig{f7}.
The observable calculated numerically  is independent of the optical potential simulating the $c$-$T$ interaction.
The black dash-dotted lines in \Fig{f7} has been obtained with a different $V_{cT}$, and although the corresponding cross sections for the individual processes---viz. elastic scattering and breakup---vary significantly from the initial calculations, their ratio is nearly superimposed to the original one.
The Ratio Method could thus provide valuable information about such $d$-bound states if it were used in its \emph{numerical} version \cite{YCP19}, i.e. by comparing experimental data to fully dynamical calculations of the reaction (CDCC or DEA).
Albeit less elegant and practical than a direct comparison to the form factor \eq{e10}, it presents the advantage to remove the dependence of the reaction calculation upon $V_{cT}$, one of its most uncertain inputs \cite{CGB04}.

\section{Conclusion}\label{conclusion}

A detailed analysis of the reaction mechanism performed in collaboration with Mahir Hussein, whose memory we honour in this review, has shown that one-neutron halo nuclei are scattered similarly whether they remain bound, i.e. when they are elastically scattered by the target, or when they break up into their more fundamental components, their core and the halo neutron \cite{CHB10}.
This result can be qualitatively understood within the Recoil Excitation and Breakup model of reaction (REB), in which the excitation of the loosely-bound projectile is due to the recoil of the core following its interaction with the target, while the neutron, seen here as a spectator, follows a mainly undisturbed path \cite{JAT97,Joh98}.
This analysis of the results of \Ref{CHB10} has led to the Ratio Method \cite{CJN11}, which suggests the study of loosely-bound nuclear structures, such as halo nuclei, by looking at the ratio of angular distributions for different processes---viz. elastic scattering and breakup.
Within the REB this ratio equals a form factor that is function of the sole projectile wave functions.
It should therefore provide an observable highly sensitive to the projectile nuclear structure by removing most of the dependence on the reaction process.

This idea has been studied by comparing the prediction of the REB to the results of accurate reaction calculations performed within the DEA \cite{BCG05,GBC06} and/or CDCC \cite{Aus87,fresco}, which include the interaction between the valence nucleon of the projectile and the target and which do not rely on the adiabatic approximation assumed in the REB.

The initial results obtained for one-neutron halo nuclei impinging on a target at intermediate energy, viz. 70~MeV/nucleon, have confirmed the validity of this idea and the minor role played at this energy and for this kind of projectile by the n-$T$ interaction and the adiabatic approximation made within the REB \cite{CJN11,CJN13}.
In those conditions a direct comparison of experimental data to the REB form factor \eq{e10} is sensible, which would both ease the analysis of such reaction measurements and help put strong constraints on the structure of the projectile.

The excellent results obtained in this original study have led us to explore the extension of the Ratio Method to low energy \cite{CCN16} and proton-rich projectiles \cite{YCP19}.
In both cases, accurate reaction calculations have confirmed the independence of the Ratio to the $c$-$T$ optical potential, which is usually poorly known far from stability.
However, a direct comparison of experimental data with the REB prediction would be less accurate because of the increasing role of $V_{{\rm N}T}$ under these conditions.
Nevertheless, when the projectile displays a clear halo structure, i.e. with a valence nucleon loosely bound to the core in a low $l$ orbital, viz. $l\le1$, and when the reaction is measured on a light target, the Ratio Method remains valid, suggesting that this new reaction observable constitutes an ideal tool to search for such exotic systems away from stability and provides detailed information about their nuclear structure.
In the case of more deeply bound systems, or a valence nucleon N bound with larger orbital angular momenta, taking $V_{{\rm N}T}$ into account, e.g., at the perturbative level, may significantly improve the REB prediction, and hence extend the Ratio Method beyond the sole realm of halo nuclei.
Since nucleon-nucleus optical potentials are rather well constrained, especially for the stable targets used at RIB facilities, such a correction should be well under control.
We hope to include such a correction within the REB and use it to improve the Ratio Method in the near future.

\section*{Acknowledgements}
PC acknowledges support from the Deutsche Forschungs\-gemeinschaft (DFG, German Research Foundation) -- Pro\-jekt-ID 279384907 -- SFB 1245 and Projekt-ID 204404729 -- SFB 1044, the European Union's Horizon 2020 research and innovation programme under Grant Agreement No. 654002, and the PRISMA+ (Precision Physics, Fundamental Interactions and Structure of Matter) Cluster of Excellence.
RCJ  acknowledges support from the UK Science and Technology Facilities Council through Grant No. STFC ST/000051/1.
FN acknowledges support from the National Science Foundation under Grant  PHY-1811815.

%

\end{document}